\documentclass{article}

\usepackage[english]{babel}
\usepackage[utf8]{inputenc}

\usepackage{amsmath}
\usepackage{mathtools}
\usepackage{amsthm}

\usepackage{amsfonts}

\usepackage[scientific-notation=true]{siunitx}

\usepackage[dvipsnames,svgnames,x11names]{xcolor}
\usepackage[marginparwidth=1.7in]{geometry}
\usepackage[markup=underlined]{changes}

\usepackage[new]{old-arrows}
\usepackage{tikz-cd}

\definechangesauthor[color=BrickRed]{IR}
\definechangesauthor[color=NavyBlue]{MZ}

\usepackage{todonotes}
\setcommentmarkup{\todo[color={authorcolor!20},size=\scriptsize]{#3: #1}}

\theoremstyle{remark}

\theoremstyle{definition}

\usepackage{float}
\usepackage{hyperref}

\title{The FRTB-IMA computational challenge for Equity Autocallables\footnote{The ideas or opinions expressed in this paper do not reflect in any way the ideas or opinions of his/her present or past employers. The techniques described in the paper, and results shown, have been obtained and are being used with the best of the authors’ knowledge and skill. No guarantee or provision is implicitly provided regarding the accuracy of the results nor any Intellectual Property rights. Some of the computational methods used in this paper fall under the scope of a number of patents; for more information, please contact Ignacio Ruiz.}}

\author{Mariano Zeron \footnote{m.zeron@mocaxintelligence.com}   \\ Meng Wu \footnote{mengwu.mw9@gmail.com}
\\ Ignacio Ruiz\footnote{i.ruiz@mocaxintelligence.com}}

\begin{document}

\maketitle

\begin{abstract}
In \cite{rz_slider_wilmott}, the Orthogonal Chebyshev Sliding Technique was introduced and applied to a portfolio of swaps and swaptions within the context of the FRTB-IMA capital calculation. The computational cost associated to the computation of the ES values – an essential component of the capital calculation under FRTB-IMA – was reduced by more than $90\%$ while passing PLA tests. 

This paper extends the use of the Orthogonal Chebyshev Sliding Technique to portfolios of equity autocallables defined over a range of spot underlyings. Results are very positive as computational reductions are of about $95\%$ with passing PLA metrics. 

Since equity autocallables are a commonly traded exotic trade type, with significant FRTB-IMA computational costs, the extension presented in this paper constitutes an important step forward in tackling the computational challenges associated to an efficient FRTB-IMA implementation.
\end{abstract}

\section{Introduction}

The capital calculation under the Fundamental Review of the Trading Book regulation (FRTB) in the Internal Models Approach (IMA), requires the daily calculation of expected shortfalls (ES) for whole range of risk factor scenarios. The regulation stipulates that each ES value must be calculated using a distribution of $250$ PnLs. Moreover, the distribution of PnLs used must satisfy the metrics specified in the PLA tests – the latter designed to make sure the PnLs used align with the PnLs generated in Front Office (daily Hypothetical PnLs). This forces financial institutions to either align or directly price their books with Front Office systems increasing computational burdens on financial institutions (see \cite{FRTB ISDA} for details).

In \cite{rz_slider_wilmott}, the Orthogonal Chebyshev Sliding Technique was applied to a portfolio of swaps and swaptions reducing pricing computational costs down by more than $90\%$, while passing PLA tests and producing small relative errors.

The aim of this paper is to apply the Orthogonal Chebyshev Sliding Technique to a portfolio of autocallables and obtain results as good as the ones in \cite{rz_slider_wilmott}.

Autocallable trades were selected as they constitute a type of exotic trade that places high computational demands on the pricing engines of financial institutions. The high cost is the result of expensive pricing routines – often relying on Monte Carlo simulations – and relatively large volumes of such trade type in several financial institutions. 

For full details of the Orthogonal Chebyshev Sliding Technique, we refer to \cite{rz_slider_wilmott} and \cite{rz_book}. This paper presents the details of how the technique is applied to the portfolio of autocallables considered and the results obtained.

At the time of writing, only a limited number of banks have announced their intention to develop the Internal Models Approach (IMA) under the FRTB regulation, despite the significantly higher capital requirements associated with the Standardised Approach (SA) compared to IMA. According to industry reports, one of the primary obstacles to the implementation of IMA is the PLA test \cite{EY Basel 3 survey}. This paper proposes a methodological and technological solution to address the challenges posed by the PLA test.

\section{Tests}\label{sec: tests}

The portfolio tested consisted of equity autocallables with a varying number of underlyings. The testing portfolio was artificially and randomly generated with some constrains, with the intention to reflect a portfolio that a financial institution could have. The portfolio consisted of several hundred Equity Autocallable trades from which around $20$\% had one single Equity underlying, $54$\% had three underlyings, $25$\% had five and $1$\% had ten.

All autocallables depend on three different types of risk factors. The first is the spot underlying. As mentioned above, every autocallable depends on either one, three, five or ten of these underlyings. The second risk factor type is volatility. There is one volatility surface per spot underlying. Therefore, an autocallable with three underlyings has three volatility surfaces that are part of the input to the pricing function of the trade. Each volatility surface had $72$ tenors points in our tests. The third risk factor is interest rates. The autocallables in the tested portfolio were sensitive to up to $3$ different interest rate curves, each with $32$ points in our tests. Taking into account the number of spot underlyings, the number of tenor points in each volatility surface and curve, there were a total of between $169$ and $826$ individual risk factors that constitute the input to the pricing functions of the autocallables tested.

For the trades considered, the FRTB-IMA capital calculation requires computing an ES value per liquidity horizon using both extended and reduced risk factor scenarios for the current period, in addition a period of stress calculation. The number of risk factor scenarios that need to be evaluated depends on the different liquidity horizons that impact the trade. For the autocallables tested, the pricing function has to be called between $1,000$ and $2,000$ times every day to be compliant with FRTB-IMA.

The PLA metrics applies to the portfolio revaluations in which all risk factors are shocked. The benchmark for the test was obtained by evaluating these scenarios using the Front Office pricing engine --– this is what constitutes the Hypothetical PnL under FRTB-IMA, distribution with respect to which the PnLs obtained with Chebyshev Sliders are measured.

As described in \cite{rz_slider_wilmott} and \cite{rz_book}, the Orthogonal Chebyshev Slider Technique involves applying Principal Component Analysis to the set of risk factor scenarios that constitute the input of the pricing functions of the trades in the portfolio. This is done to reduce the dimensionality of the pricing function input space, which in turn allows for the building of Chebyshev Sliders in a computational cheap manner without significant loss of accuracy. 

As mentioned above, the autocallables considered have three risk factor types. PCA was applied to each risk factor type \emph{independently} of the others. The reason why this is done is two-fold. The first is that the correlation between risk factors that belong to the same risk factor type is high – something that is usually not the case for risk factors belonging to different types. The second is that by building a PCA object for each risk factor type, the Chebyshev Slider can be used for the IMA calculations that involve partial shocks of risk factors, without having to re-calibrate the Chebyshev Slider (see \cite{rz_slider_wilmott} or \cite{rz_book} for a full description of why this is the case). Therefore, in the tests performed, a total of three PCA objects were built: one for Equity spots, one for Equity volatilities and one for interest rates.

In the case of the spot underlyings, all principal components were used in the calculation. The reason is that spot underlyings typically display a relatively low level of correlation. Therefore, one cannot reduce the dimension of the space aggressively without losing a sizeable amount of information. Additionally, the largest number of spot underlyings in the test was ten, which does not constitute a large number of dimensions for a Chebyshev Slider, hence reducing the number of components coming from the spot underlyings would not have had any significant effect in the final solution. It is important to note, however, that one may consider reducing the dimension of the space of spot underlyings when there are many of them and if their correlation is high.

In the case of volatilities, the first $3$ principal components were kept while the rest were discarded. In the case of interest rates the first $6$ components were retained As results in the next section show, this number of components were enough to pass the PLA.

As discussed in \cite{rz_slider_wilmott} and \cite{rz_book}, the dimension of the domain of any Chebyshev Slider is the sum of the principal components kept after applying PCA to all risk factor types. Therefore, the dimension of the slider ranged from $10$ in the case of the autocallables with single underlying, to $19$ in the case of autocallables with ten underlyings.

A Chebyshev Slider was built per trade. The configuration of the sliders selected consisted of slides of dimension one, each with a total of $6$ Chebyshev points.  This means the sliders had as few as $61$ Chebyshev calibrating scenarios in the case of single underlying autocallables, to $115$ calibrating scenarios in the cases of ten underlyings.

Each of the Chebyshev Sliders was evaluated on all sets of $250$ scenarios that are needed for the capital calculation as well as on the period of stress scenarios; i.e. all liquidity horizons involved, partial shocks and required period of stressed calculations. In addition to the PLA tests, we measure the relative ES error. These results are presented in the following section.

\section{Results}\label{sec: results}

Computational gains obtained when using Chebyshev Sliders are estimated from the number of pricing calls to the risk engine required to build the Chebyshev Sliders and the number of calls needed to obtain all the ES values needed for the capital calculation as well as the period of stress calculation. There are the two reasons that justify the previous assertion.

The first is that the slider built can be used to obtain PnLs for all liquidity horizons needed under FRTB-IMA and not just the PnLs that correspond to the $10$-day liquidity horizon. The second is that the speed of evaluation of the Chebyshev Sliders compared to that of the pricing routines of autocallables is significant enough that obtaining the PnLs required for all ES values is miniscule compared to the time it takes to build the tensors (see \cite{rz_slider_wilmott} and \cite{rz_book} for more details on this point).

As mentioned above, the number of Chebyshev scenarios varied depending on the number of autocallable underlyings. For one underlying there were $61$, for autocallables with $3$ underlyings $73$, in the case of $5$ underlyings $85$, and $115$ in the case of $10$ underlyings. If we assume $80$ Chebyshev scenarios on average, then the building cost of the Chebyshev Sliders is essentially the cost of pricing the portfolio of autocallables $80$ times. Given that we need around $1,500$ calls to the pricing function of autocallables to run the FRTB-IMA calculation, this gives a computational reduction of about $95$\%.

\begin{figure}[H]
\centering
\includegraphics[scale=0.85]{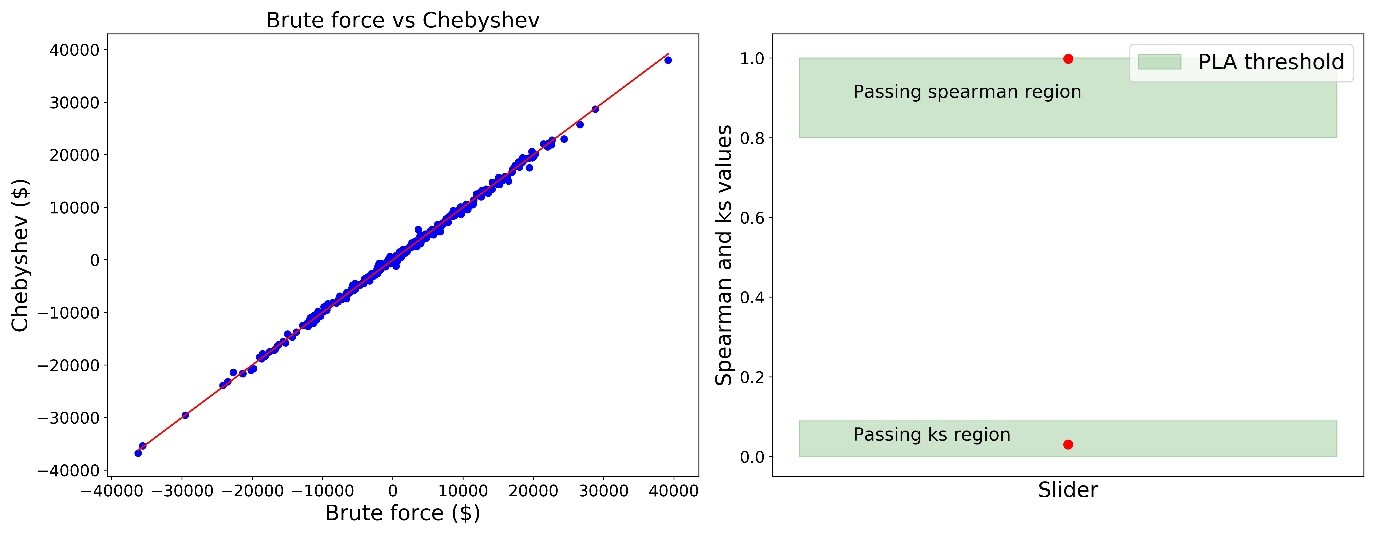}
\caption{Left hand plot compares the PnLs obtained with the original pricing model with the PnLs obtained with the Chebyshev slider on a single autocallable evaluated on . The plot on the right shows the Spearman correlation and the kolmogorov smirnov value between such PnLs.}
\label{fig: single auto}
\end{figure}

\begin{figure}[H]
\centering
\includegraphics[scale=0.85]{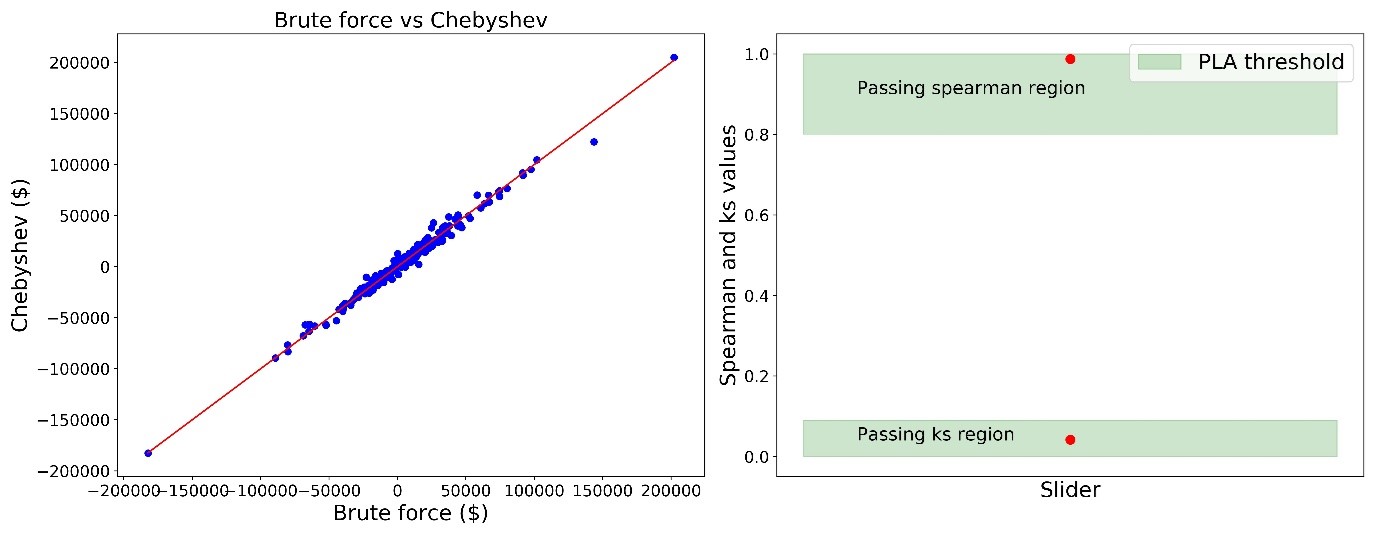}
\caption{Left hand plot compares the PnLs obtained with the original pricing model with the PnLs obtained with the Chebyshev slider on the portfolio of autocallables evaluated on . The plot on the right shows the Spearman correlation and the kolmogorov smirnov value between such PnLs.}
\label{fig: portfolio auto}
\end{figure}

\begin{figure}[H]
\centering
\includegraphics[scale=0.85]{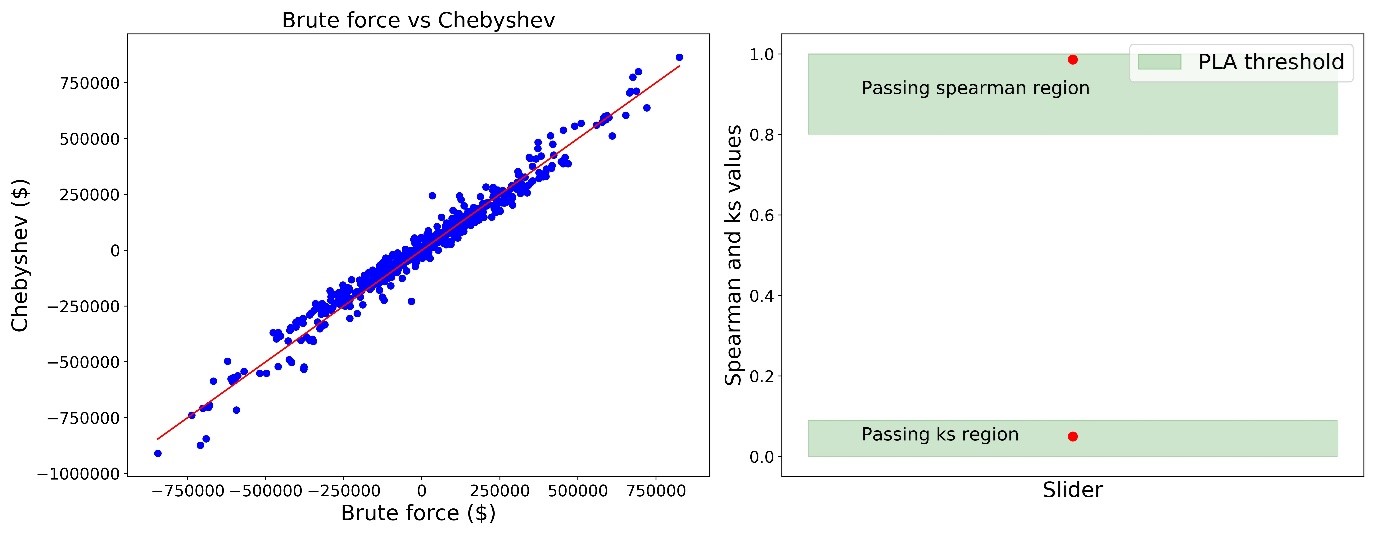}
\caption{Left hand plot compares the PnLs obtained with the original pricing model with the PnLs obtained with the Chebyshev slider on the portfolio of autocallables evaluated on the period of stress scenarios. The plot on the right shows the Spearman correlation and the kolmogorov smirnov value between such PnLs.}
\label{fig: portfolio auto stress}
\end{figure}

The accuracy of the results in terms of the PLA test is shown in Figures \ref{fig: single auto}, \ref{fig: portfolio auto} and \ref{fig: portfolio auto stress}. In each of them, the plot on the left-hand side shows a comparison of the PnLs obtained with the risk system compared to the PnLs obtained using Chebyshev Sliders. The plot on the right shows the values of both the Spearman correlation and the KS value (red dots) obtained from the two sets of PnLs on the left. The same graph also shows the areas (in green) that correspond to passing the PLA tests for the proxy pricer. In addition to the PLA metrics, the ES value was computed using both sets of PnLs.

Figure \ref{fig: single auto} shows the result obtained on a typical single autocallable with $10$ spot underlyings – the greatest number of underlyings present in the portfolio. Figure \ref{fig: portfolio auto} shows results on the whole portfolio of autocallables evaluated on the scenarios that correspond to the $10$-day liquidity horizon (i.e. all risk factors are shocked). Figure \ref{fig: portfolio auto stress} shows the results for the whole portfolio of autocallables evaluated on the period of stress scenarios. In every single case, the PLA metrics obtained are well within the range that is needed to pass the PLA test.

To be noted, the computational reduction just estimated assumes daily Chebyshev Slider calibration. However, if the Chebyshev Slider is used more than once, the total computational savings are greater. For example, Chebyshev Slider could be built during the weekend and used throughout the week. In this approach, sliders would need to be built for incoming trades, but the bulk of the portfolio would remain unchanged, reducing the daily FRTB-IMA calculation to essentially the evaluation of already existing Chebyshev Sliders. Not only would the daily FRTB-IMA calculation become more affordable to run but this can open the door to a range of intra-day calculations – for example, what-if type of calculations – that are currently not possible due to lack of computational power.

\section{Conclusions}\label{sec: conclusion}

The Orthogonal Chebyshev Sliding Technique (see \cite{rz_slider_wilmott} and \cite{rz_book}) was applied on a portfolio of autocallables within the context of the FRTB-IMA regulation. Computational cost reductions of about $95$\% were obtained while passing the PLA tests.

This constitutes an important step forward in the attempt to reduce the computational burden associated FRTB-IMA since the autocallable is an exotic trade with costly pricing routine that is commonly present in the portfolios of financial institutions.

The results shown in this paper are limited to only one trade type (equity autocallables) for simplicity of illustration. This trade type was chosen because it is considered a very exotic payoff (hence difficult to replicate via proxy functions) that has a substantial computational evaluation cost. However, it should be noted that the authors have applied this technique to a range of typical trade types, both vanilla and exotic, in different asset classes, with similar positive results. The evidence presented in \cite{rz_slider_wilmott}, \cite{rz_book} and this paper suggest that the Orthogonal Chebyshev Sliding Technique can be extended to several other trade types with similar results.

The authors can think of two avenues for future research. A key question is how often to run the calibration step. This paper assumes daily calibration of the Chebyshev Sliders. However, one could think of weekly, monthly or even less frequent calibrations, with the benefit that the computational cost improvement could go then well beyond $99$\%. In the authors’ view, this could be achieved by introducing an extra slide in the Chebyshev Slider that accounts for the change in value of the derivative over time (similarly to the Theta Greek in a Taylor type expansion). Additionally, other forms of data dimensionality reduction, such as Variational Autoencoders and Differential PCA (\cite{huge_savine_pca}), could also be studied to improve performance.


\begin{thebibliography}{1}





\bibitem{rz_slider_wilmott} Ruiz, I. Zeron, M. 2021. Denting the FRTB-IMA Computational Challenge via Orthogonal Chebyshev Sliding Technique. \emph{Wilmott}


\bibitem{rz_book} Ruiz, I. Zeron, M. 2021. 2021 \emph{Machine Learning for Risk Calculations: A Practitioner's View}. Wiley.

\bibitem{huge_savine_pca} Huge, B. Savine, A. 2021. Axes that matter: PCA with a difference. \emph{Risk}. 

\bibitem{FRTB ISDA} Basel Committee on Banking Supervision. 2024. MAR Calculation of RWA for market risk. BIS.

\bibitem{EY Basel 3 survey} EY. 2023 Global Basel 3 Reforms survey.

\end{thebibliography}
\end{document}